\documentclass[
 amsmath,amssymb,
 aps,
]{revtex4-2}
\usepackage[utf8]{inputenc}
\usepackage{graphicx}
\usepackage{dcolumn}
\usepackage{hyperref}

\begin{document}

\preprint{}

\title{Comment on "ENN's roadmap for proton-boron fusion based on spherical torus" [Phys. Plasmas 31, 062507 (2024)]}

\author{Zhi Li}
\email{zli092902@gmail.com}
\address{Washington, DC 20005, United States}

\date{June 30, 2024}

\begin{abstract}
This comment discusses the feasibility of hot ion mode ${{T}_{i}}/{{T}_{e}}=4$ for proton-boron fusion which is critical for the roadmap proposed in Liu, M et al [Phys. Plasmas 31, 062507 (2024)]. The hot ion mode ${{T}_{i}}/{{T}_{e}}=4$ has been calculated to be far from accessible (${{T}_{i}}/{{T}_{e}}<1.5$ for ${{T}_{i}}=150\text{keV}$) under the most optimal conditions if fusion provides the heating (Xie, H [Introduction to Fusion Ignition Principles: Zeroth Order Factors of Fusion Energy Research (in Chinese), USTC Press, Hefei, 2023]), i.e., that all fusion power serves to heat the ions and that electrons acquire energy only through interactions with ions. We also discuss if hot ion mode of ${{T}_{i}}/{{T}_{e}}=4$ could be achieved by an ideal heating method which is much more efficient than fusion itself (near 20 times fusion power for ${{T}_{i}}=150\text{keV}$) and only heats the ions, whether it makes sense economically.  
\end{abstract}

\maketitle
Nuclear fusion with advanced fuels has always been conjectured to be the final resolution for commercial fusion, among which proton-boron fusion is attracting growing attention in recent years. Operating in hot ion mode is one of the proposed approaches to commercial proton-boron fusion.

As was discussed in \cite{xie}, hot ion mode with high temperature difference between electrons and ions of proton-boron fusion is hard to maintain to the extent that even if all fusion power is converted to heat ions and electrons only acquire energy through their interactions with ions, for ${{T}_{i}}=150\text{keV}$, the proposed operating temperature for proton-boron fusion, ${{T}_{i}}/{{T}_{e}}<1.5$. And this ratio decreases as ${{T}_{i}}$ decreases.

Here we just translate the content on page 49 of \cite{xie}. 

"We make the simplest calculation to estimate the electron temperature in the hot ion mode of proton-boron fusion.

\begin{figure}[htbp] 
\centering
\includegraphics[width=3.5in]{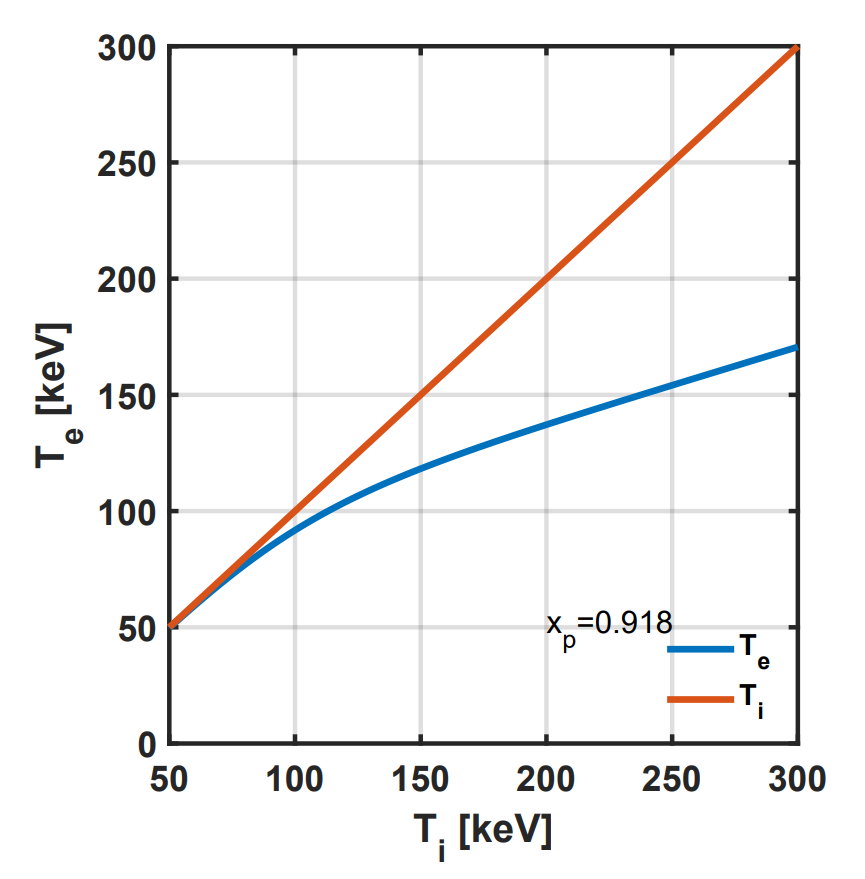} 
\caption{The relationship between ${T}_{i}$ and ${T}_{e}$ for hot ion mode of proton-boron fusion. Under the most optimal condition, from Figure 3.17(a) in \cite{xie} }\label{fig:1}  
\end{figure}

Assume that all the power produced by proton-boron fusion serves to heat ions, and the electrons get energy only through thermal exchange with ions. Neglecting the loss due to radiation and transport, we only do the estimation to the leading order

\begin{equation}\label{energy}
    -\frac{d{{W}_{i}}}{dt}=\frac{3}{2}{{k}_{B}}\left( \frac{{{n}_{p}}}{{{\tau }_{p,e}}}+\frac{{{n}_{B}}}{{{\tau }_{B,e}}} \right)({{T}_{i}}-{{T}_{e}})={{f}_{ion}}{{P}_{fus}}
\end{equation}
(which is (3.62) in \cite{xie}. And ${{P}_{fus}}$ is fusion power, ${{n}_{p}}$ and ${{n}_{B}}$ are the densities of proton and boron ions respectively, ${{f}_{ion}}={{Y}_{+}}/Y$ is the proportion of ions in all energy carrying particles which is assumed to be 1 in \cite{xie}, and $Y$ is the energy emitted by a single nuclear reaction, ${{Y}_{+}}$ is the energy of charged particles in a single nuclear reaction; ${{T}_{i}}$ and ${{T}_{e}}$ are the temperatures of ions and electrons respectively)
where ${{\tau }_{p,e}}$ and ${{\tau }_{B,e}}$ are the thermal exchange times of proton and boron ions respectively, which can be calculated by eq \ref{3.48} (equation 3.48 in \cite{xie}), and the result is depicted in Figure \ref{fig:1}. The electron temperature thus calculated is the minimum possible steady state temperature. Taking other effects into consideration, the practical electron temperature will be larger. It can be seen that for ${{T}_{i}}=300\text{keV}$, ${{T}_{e}}=170\text{keV}$, indicating ${{T}_{e}}/{{T}_{i}}=0.57>0.5$. For smaller ${{T}_{i}}$, the difference between ${{T}_{i}}$ and ${{T}_{e}}$ is even smaller. This demonstrates that it is very hard to maintain hot ion mode with large temperature difference between ${{T}_{i}}$ and ${{T}_{e}}$. This conclusion is similar to \cite{moreau1977potentiality} and \cite{dawson}. "

\begin{equation}\label{3.48}
    {{\tau }_{ij}}=\frac{3\sqrt{2}{{\pi }^{3/2}}\varepsilon _{0}^{2}{{m}_{i}}{{m}_{j}}}{{{n}_{j}}{{e}^{4}}Z_{i}^{2}Z_{j}^{2}\ln \Lambda }{{\left( \frac{{{k}_{B}}{{T}_{i}}}{{{m}_{i}}}+\frac{{{k}_{B}}{{T}_{j}}}{{{m}_{j}}} \right)}^{3/2}}
\end{equation}

\begin{figure}[htbp] 
\centering
\includegraphics[width=3.5in]{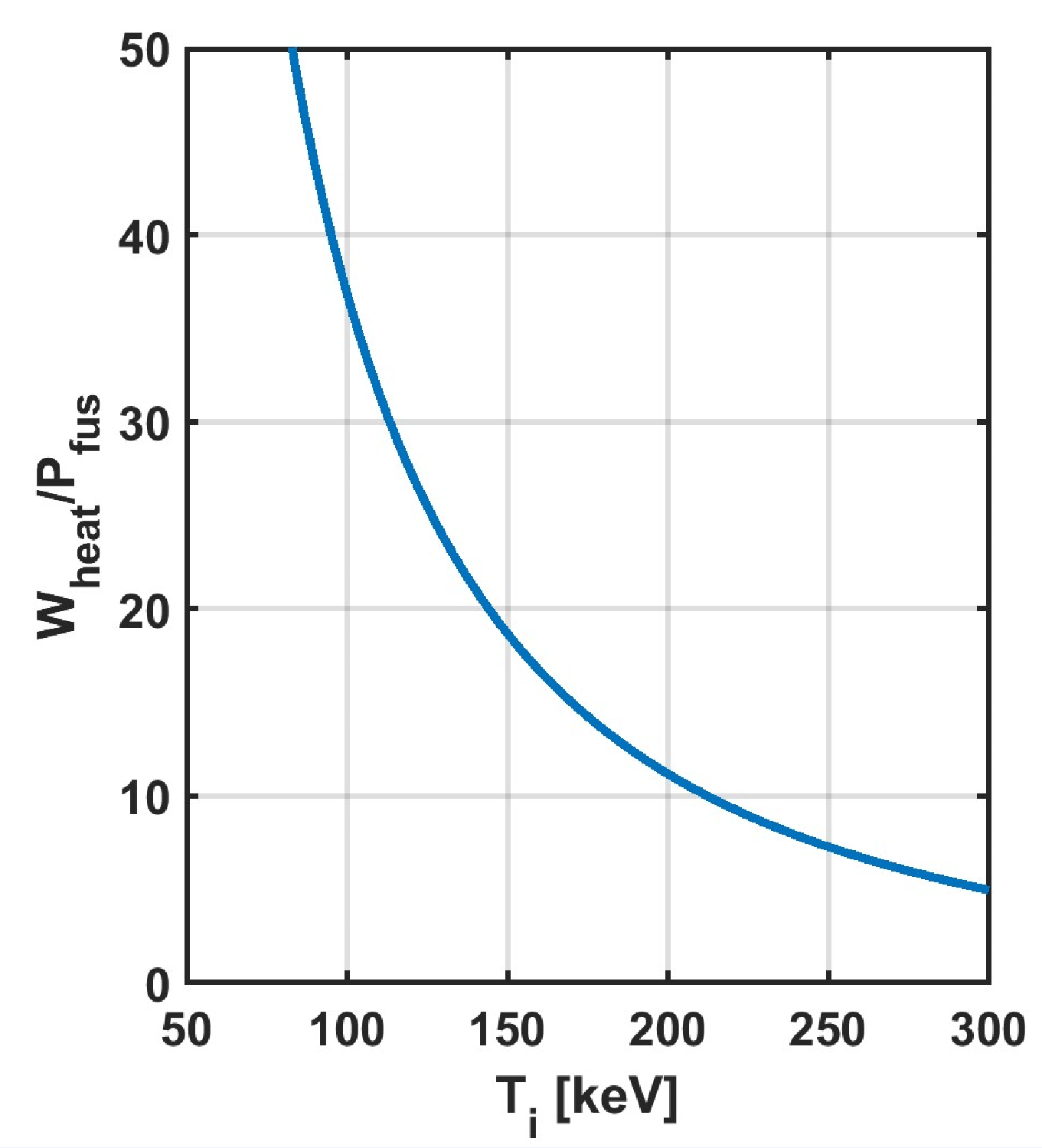} 
\caption{The ratio between heating power and fusion power for ${{T}_{i}}/{{T}_{e}}=4$}\label{fig:2}  
\end{figure}

Now let us assume that there is an ideal heating method, to the extreme that it only heats ion, and that it leads to no additional energy loss. This idealized heating enables hot ion mode of ${{T}_{i}}/{{T}_{e}}=4$. Denote it by ${{W}_{heat}}$, the balance equation \ref{energy} transforms into

\begin{equation}\label{power}
    -\frac{d{{W}_{i}}}{dt}=\frac{3}{2}{{k}_{B}}\left( \frac{{{n}_{p}}}{{{\tau }_{p,e}}}+\frac{{{n}_{B}}}{{{\tau }_{B,e}}} \right)({{T}_{i}}-{{T}_{e}})={{f}_{ion}}{{P}_{fus}}+{{W}_{heat}}
\end{equation}
eq\ref{power} actually becomes a linear function for ${{W}_{heat}}$. Adapting the code from  \href{http://hsxie.me/fusionbook/}{http://hsxie.me/fusionbook/}, we can easily calculate ${{W}_{heat}}$ and compare it with the fusion power ${{f}_{ion}}{{P}_{fus}}$. The result is plotted in figure \ref{fig:2}. 

It is apparent that even the idealized $W_{heat}$ that enables ${{T}_{i}}/{{T}_{e}}=4$ is way way larger than fusion power.

\begin{figure}[htbp] 
\centering
\includegraphics[width=3.5in]{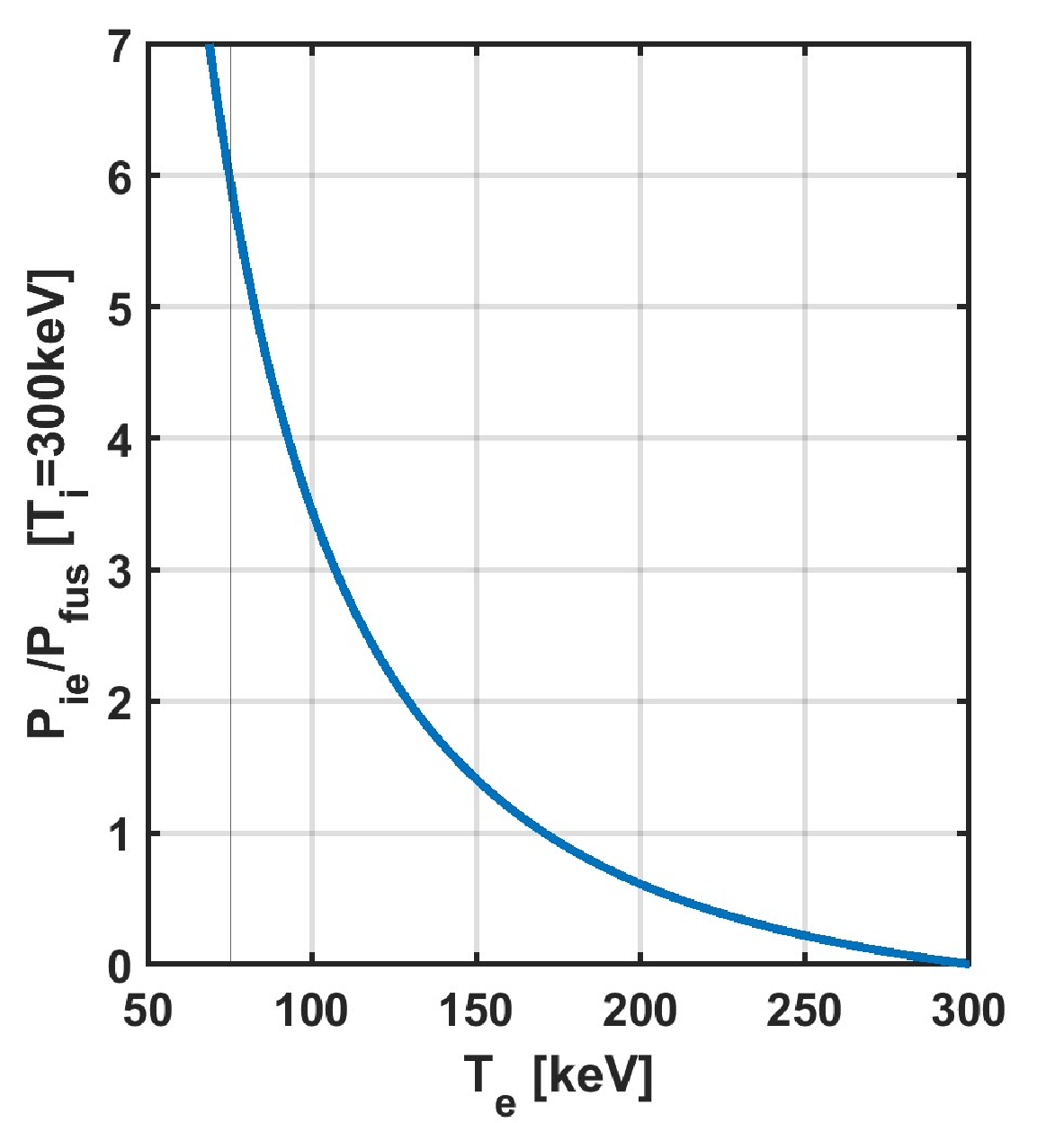} 
\caption{The power of energy exchange between electrons and ions vs fusion power for ${{T}_{i}}=300\text{keV}$. Adapted from Figure 3.17(b) of \cite{xie}}\label{fig:3}  
\end{figure}

In fact, after slight extension of the axis in Figure 3.17(b)  of \cite{xie}, i.e.,\ref{fig:3}, it is already clear that for ${{T}_{i}}=300\text{keV}$, hot ion mode of ${{T}_{i}}/{{T}_{e}}=4$ demands a heating that is about 5 times the fusion power.

The situation is therefore very awkward: if electron temperature is high, Bremsstrahlung radiation is significant; and if the electron temperature is kept to be low, the power to sustain hot ion mode would be much larger than fusion power. Now imagine that we have finally succeeded in finding such an ideal heating method, a question emerges naturally: why don't we just use its power than bother to do fusion?

This analysis also leads to a question on the validity of the scaling law for the energy confinement time ${{\tau }_{E}}$ of proton-boron fusion\cite{liu2024}. Were it rigorous, the fusion energy gain factor $Q$ calculated from the triproduct $nT{{\tau }_{E}}$ or the power balance with term $\frac{{{W}_{th}}}{{{\tau }_{E}}}$ would be large\cite{liu2024}. But if we calculate $Q$ directly, then it is vastly smaller. And if we were not lucky enough to find such an ideal heating, ${{\tau }_{E}}$ does not exist at all for proton-boron fusion with ${{T}_{i}}/{{T}_{e}}=4$.

\acknowledgments
We thank Dr Xie for the permission to reproduce the figures in his book.

\bibliography{apssamp}

\providecommand{\noopsort}[1]{}\providecommand{\singleletter}[1]{#1}%
\begin{thebibliography}{4}%
\makeatletter
\providecommand \@ifxundefined [1]{%
 \@ifx{#1\undefined}
}%
\providecommand \@ifnum [1]{%
 \ifnum #1\expandafter \@firstoftwo
 \else \expandafter \@secondoftwo
 \fi
}%
\providecommand \@ifx [1]{%
 \ifx #1\expandafter \@firstoftwo
 \else \expandafter \@secondoftwo
 \fi
}%
\providecommand \natexlab [1]{#1}%
\providecommand \enquote  [1]{``#1''}%
\providecommand \bibnamefont  [1]{#1}%
\providecommand \bibfnamefont [1]{#1}%
\providecommand \citenamefont [1]{#1}%
\providecommand \href@noop [0]{\@secondoftwo}%
\providecommand \href [0]{\begingroup \@sanitize@url \@href}%
\providecommand \@href[1]{\@@startlink{#1}\@@href}%
\providecommand \@@href[1]{\endgroup#1\@@endlink}%
\providecommand \@sanitize@url [0]{\catcode `\\12\catcode `\$12\catcode `\&12\catcode `\#12\catcode `\^12\catcode `\_12\catcode `\%12\relax}%
\providecommand \@@startlink[1]{}%
\providecommand \@@endlink[0]{}%
\providecommand \url  [0]{\begingroup\@sanitize@url \@url }%
\providecommand \@url [1]{\endgroup\@href {#1}{\urlprefix }}%
\providecommand \urlprefix  [0]{URL }%
\providecommand \Eprint [0]{\href }%
\providecommand \doibase [0]{https://doi.org/}%
\providecommand \selectlanguage [0]{\@gobble}%
\providecommand \bibinfo  [0]{\@secondoftwo}%
\providecommand \bibfield  [0]{\@secondoftwo}%
\providecommand \translation [1]{[#1]}%
\providecommand \BibitemOpen [0]{}%
\providecommand \bibitemStop [0]{}%
\providecommand \bibitemNoStop [0]{.\EOS\space}%
\providecommand \EOS [0]{\spacefactor3000\relax}%
\providecommand \BibitemShut  [1]{\csname bibitem#1\endcsname}%
\let\auto@bib@innerbib\@empty
\bibitem [{\citenamefont {Xie}(2023)}]{xie}%
  \BibitemOpen
  \bibfield  {author} {\bibinfo {author} {\bibfnamefont {H.}~\bibnamefont {Xie}},\ }\href@noop {} {\emph {\bibinfo {title} {Introduction to Fusion Ignition Principles: Zeroth Order Factors of Fusion Energy Research (in Chinese)}}}\ (\bibinfo  {publisher} {USTC Press},\ \bibinfo {year} {2023})\BibitemShut {NoStop}%
\bibitem [{\citenamefont {Moreau}(1977)}]{moreau1977potentiality}%
  \BibitemOpen
  \bibfield  {author} {\bibinfo {author} {\bibfnamefont {D.~C.}\ \bibnamefont {Moreau}},\ }\bibfield  {title} {\bibinfo {title} {Potentiality of the proton-boron fuel for controlled thermonuclear fusion},\ }\href@noop {} {\bibfield  {journal} {\bibinfo  {journal} {Nuclear Fusion}\ }\textbf {\bibinfo {volume} {17}},\ \bibinfo {pages} {13} (\bibinfo {year} {1977})}\BibitemShut {NoStop}%
\bibitem [{\citenamefont {Teller}(1981)}]{dawson}%
  \BibitemOpen
  \bibfield  {author} {\bibinfo {author} {\bibfnamefont {E.}~\bibnamefont {Teller}},\ }\href@noop {} {\emph {\bibinfo {title} {Fusion}}}\ (\bibinfo  {publisher} {New York: Academic},\ \bibinfo {year} {1981})\BibitemShut {NoStop}%
\bibitem [{\citenamefont {Liu}\ \emph {et~al.}(2024)\citenamefont {Liu}, \citenamefont {Xie}, \citenamefont {Wang}, \citenamefont {Dong}, \citenamefont {Feng}, \citenamefont {Gu}, \citenamefont {Huang}, \citenamefont {Jiang}, \citenamefont {Li}, \citenamefont {Li} \emph {et~al.}}]{liu2024}%
  \BibitemOpen
  \bibfield  {author} {\bibinfo {author} {\bibfnamefont {M.-s.}\ \bibnamefont {Liu}}, \bibinfo {author} {\bibfnamefont {H.-s.}\ \bibnamefont {Xie}}, \bibinfo {author} {\bibfnamefont {Y.-m.}\ \bibnamefont {Wang}}, \bibinfo {author} {\bibfnamefont {J.-q.}\ \bibnamefont {Dong}}, \bibinfo {author} {\bibfnamefont {K.-m.}\ \bibnamefont {Feng}}, \bibinfo {author} {\bibfnamefont {X.}~\bibnamefont {Gu}}, \bibinfo {author} {\bibfnamefont {X.-l.}\ \bibnamefont {Huang}}, \bibinfo {author} {\bibfnamefont {X.-c.}\ \bibnamefont {Jiang}}, \bibinfo {author} {\bibfnamefont {Y.-y.}\ \bibnamefont {Li}}, \bibinfo {author} {\bibfnamefont {Z.}~\bibnamefont {Li}}, \emph {et~al.},\ }\bibfield  {title} {\bibinfo {title} {Enn's roadmap for proton-boron fusion based on spherical torus},\ }\href@noop {} {\bibfield  {journal} {\bibinfo  {journal} {Physics of Plasmas}\ }\textbf {\bibinfo {volume} {31}} (\bibinfo {year} {2024})}\BibitemShut {NoStop}%
\end{thebibliography}%

\end{document}